\documentclass[a4paper,10pt]{article}
\usepackage{amsmath}
\usepackage[pdftex]{graphicx}
\usepackage{subfig}
\usepackage[super]{natbib}

\title{YFitter: Maximum likelihood assignment of Y chromosome
  haplogroups from low-coverage sequence data}
\author{Luke Jostins, Yali Xu, Shane McCarthy, Qasim Ayub, \\
  Richard Durbin, Jeff Barrett, Chris Tyler-Smith}
\begin{document}

\maketitle

\begin{abstract}
Low-coverage short-read resequencing experiments have the potential to
expand our understanding of Y chromosome haplogroups. However, the
uncertainty associated with these experiments mean that haplogroups
must be assigned probabilistically to avoid false inferences. We propose
an efficient dynamic programming algorithm that can assign haplogroups
by maximum likelihood, and represent the uncertainty in assignment. We
apply this to both genotype and low-coverage sequencing data, and show
that it can assign haplogroups accurately and with high
resolution. The method is implemented as the program YFitter, which
can be downloaded from http://sourceforge.net/projects/yfitter/
\end{abstract}

\section{Introduction}

Low-coverage, short read resequencing is a cost effective means of
carrying out variant discovery, disease association and population
genetics experiments\citep{durbin2010}\citep{li2011}.  One potential
value of low-coverage sequencing experiments is that, as the
experiments are whole-genome, previously less well-studied regions of
the genome such as the Y chromosome are sequenced ``for free'' . Many
Y chromosome mutations have been discovered\citep{karafet2008}, and the haplogroups they define have
been found to be associated with various population genetic and
disease associations\citep{sezgin2009}\citep{underhill2009}. Large,
low coverage sequencing projects have the potential to greatly refine
our understanding of these Y haplogroups.

However, the uncertainty associated with the lower coverage and higher
error rate of these experiments has to be handled
statistically to avoid biases, and this is especially true of the Y
chromosome, due to the lower sequence coverage.. Missing or uncertain
data can result in incorrect haplogroup assignment, especially if
present high up in the haplogroup tree, which can then lead to false
inferences. In addition, assigning samples by hand in large sequencing
experiments can be time consuming, so an automated solution is
required. While probabilistic, automatic methods have been developed
for Short-Tandem Repeats (Y-STRs)\citep{athey2006}, no equivalent
method has been developed for low coverage sequencing.

The large number of different Y haplogroups in a given tree makes
separate calculation of the likelihood of sequencing reads given each
Y haplogroup computationally expensive. We propose an efficient
dynamic programming algorithm to calculate the likelihood, and use
this to assign maximum-likelihood haplogroups robustly. This method
can accurately assign individuals to haplogroups given either genotyping chip or
low-coverage sequence data, and can calculate confidence haplogroups that
take into account uncertainty using the Akaike information criteria.

We have implemented this method in C++ as the program YFitter, which is open
source and freely available.

\section{The Method}

As there is no homologous recombination on the Y chromosome, Y
haplogroups lie on a tree with each node $i$ being defined by one or
more mutations $B_{i}$. Our aim is to select the Y chromosome haplogroup that
maximises the likelihood of the observed reads, along with a lower
resolution confidence haplogroup that encompases all plausible
haplogroups.  We will do this by defining the likelihood in terms of
recursively calculable statistics, using the tree structure of the
haplogroups.

We will write $B^{+}_{i}$ if a set of mutations is present in the
individual under consideration, and $B^{-}_{i}$ if it is not. The
raw data for the algorithm is the per-site likelihood of observing
reads at the mutation sites $D_{i}$ given that the mutations $B_{i}$
have or have not occurred:

\begin{eqnarray}
M^{+}_{i} &=& P(D_{i} | B^{+}_{i}) \\
M^{-}_{i} &=& P(D_{i} | B^{-}_{i})
\end{eqnarray}

Many methods have been developed to calculate such likelihoods from
short read data \citep{depristo2011}\citep{li2009}. 

We will define all nodes downstream of node $i$ as $i\downarrow$,
and the reads and mutations at these sites $D_{i\downarrow}$ and $B_{i\downarrow}$. We can then define the
downward likelihood of node $i$ as

\begin{eqnarray}
L^{\downarrow}_{i} &=& P(D_{i\downarrow} | B^{-}_{i}, B^{-}_{i\downarrow}) \\
&=& \left\{ \begin{array}{ll}
      M^{-}_{i} & \mbox{for leaf nodes} \\
      M^{-}_{i} \prod_{j \in d(i)} L^{\downarrow}_j & \mbox{for non-leaf nodes} \end{array} \right.
\end{eqnarray}

Where $d_{i}$ is the set of daughters of node $i$. This is the
likelihood of observing reads downstream of $i$, given that the
individual's haplogroup assignment is not descended from $i$. We
calculate this starting at the leaf nodes, and work upwards to the
root node.

For each branch, we can then define the upwards likelihood:

\begin{eqnarray}
L^{\uparrow}_{i} &=& P(D_{\backslash i\downarrow} | B^{+}_{i}, B^{+}_{i\uparrow},
B^{-}_{s(i)},  B^{-}_{s(i)\downarrow}, B^{-}_{c(i)}) \\
&=& \left\{ \begin{array}{ll}
      M^{+}_{i} & \mbox{for the root node} \\
      L^{\uparrow}_{p(i)} M^{+}_{i} \prod_{j \in s(i)} L^{\downarrow}_{j} & \mbox{for non-root nodes} \end{array} \right. 
\end{eqnarray}

Where $p(i)$ is the parent of node $i$, $s(i)$ is the set of siblings of $i$,
$i\uparrow$ is the set of direct ancestors of $i$ and $c(i)$ (for
``cousins'') is the set of all nodes that are not descended from or
direct ancestors of $p(i)$.

The upwards likelihood is thus the likelihood of observing reads at
mutation sites not descended from $i$, given that the individual's
haplogroup is descended from $i$. This value is calculated from the
root node, working down to the leaf nodes.

For leaf nodes, the full likelihood is equal to the upwards
likelihood, as there is no data downstream of them. For non-leaf
haplogroups, we define the likelihood as the maximum of the likelihoods of its descendants.

\begin{eqnarray}
L_{i} &=& P(D | B^{+}_{i}, B^{+}_{i\uparrow},
B^{+}_{i\downarrow}, B^{-}_{s(i)},  B^{-}_{s(i)\downarrow}, B^{-}_{c(i)}) \\
&=& \left\{ \begin{array}{ll}
      L^{\uparrow}_{i} & \mbox{for leaf nodes} \\
      max (L_{j}: j \in d(i)) & \mbox{for non-leaf nodes} \end{array} \right. 
\end{eqnarray}

As it is likely that multiple haplogroups will all have the maximum
likelihood, the maximum likelihood haplogroup is then defined as the
haplogroup with the maximum likelihood that is closest to the
root node. This is equivalent to the most recent common ancestor of all
haplogroups with the maximum likelihood.

As well as a maximum-likelihood haplogroup, we define a haplogroup
confidence set as all haplogroups with a likelihoods within 8.685
phred-scaled log units of the maximum likelihood, equivalent to a
$\Delta$AIC of 4\citep{burnham2002}. The confidence haplogroup is thus the most recent
common ancestor of all haplogroups in this confidence set, and all
haplogroups not derived from the confidence haplogroup are judged to
have ``considerably less support''\citep{burnham2002}.

The YFitter program reads in a haplogroup tree in phyloXML
format\citep{han2009}, with mutations specified as properties of
branches. We constructed such a haplogroup tree using the mutations
catalogued by Karafet \textit{et al}\citep{karafet2008}. We removed
G/C and A/T SNPs to avoid stranding errors, as well as indels, repeats
and non-uniquely mapped variants. 

\section{Applications}

\subsection{Assigning Haplogroups to Genotype Data}

We tested our method on publicly available data from the 9 males of the Genomes Unzipped
project. All participants in the project have released genotyped data
generated by the personal genomics company 23andMe. This is a good
test set for haplogrouping, as the custom 23andMe chip is designed to
contain a large number of haplogroup-informative variants. 

We assigned haplotypes to the 9 indiviudals using our YFitter program,
and compared the results to the assignments made by
23andMe (Table 1). The set contains 3 different major haplogroups. All
individuals have broadly consistent assignments, though there is some
ambiguity between the haplogroup names within haplogroup R1b1b2 between 23andMe
(who use the ISOGG2010 tree) and Karafet \textit{et al} (the YCC2008 tree).

\begin{table}
\begin{tabular}{| l | l | l | }
\hline
Individual & YFitter haplogroup & 23andMe haplogroup \\
\hline
CAA001 & R1a1 & R1a1a* \\ 
DBV001 & J2 & J2 \\
DFC001 & R1b1b2g & R1b1b2a1a1d*\textdagger \\
DGM001 & R1a1 & R1a1a* \\
JCB001 & R1b1b2 & R1b1b2a1 \\
JKP001 & R1b1b2 & R1b1b2a1a2f \\
JXA001 & R1b1b2g & R1b1b2a1a1*\textdagger \\
LXJ001 & N1c1 & N1c1 \\
VXP001 & R1b1b2d & R1b1b2a1 \textdagger \\
\hline
\end{tabular}
\caption{Haplogroup assignments using YFitter, compared to 23andMe's
  assignments, for the Genomes Unzipped males. Variants with
  inconsistent nomenclature are marked with a \textdagger.}
\end{table}

\subsection{Assigning Haplogroups to Low Coverage Sequencing Data}

We also tested our method using publicly available data from 286 individuals present both in the Phase
I of the 1000 Genomes Project\citep{durbin2010} and Phase 3 of the
HapMap project\citep{altshuler2010}. Haplogroups were assigned
manually using the HapMap genotyping data, and automatically using
YFitter on the low-coverage 1000 Genomes Project sequencing data. Genotype
likelihoods were generated from sequence data using the program
samtools\citep{li2009}. The 286 individuals contained 12 different
major haplogroups.

Of the 286 maximum likelihood haplogroup assignments, 285 were fully consistent between
the genotype and sequence data. Of those, 203 assignments had greater
resolution in the sequence data, 71 had the same resolution, and in 11 had a lower resolution. If the
confidence haplogroup was used, there were no inconsistencies, 199 had
higher resolution, 75 had the same resolution, and 12 had lower
resolution. Both sets sequenced-based haplogroup assignments were of higher
resolution than the genotype-based set, and the confidence haplogroups
were only of slightly lower resolution than the maximum likelihood haplogroups (Figure 1).

\begin{figure}[ht]
 \centering
 \includegraphics[width=0.7\textwidth]{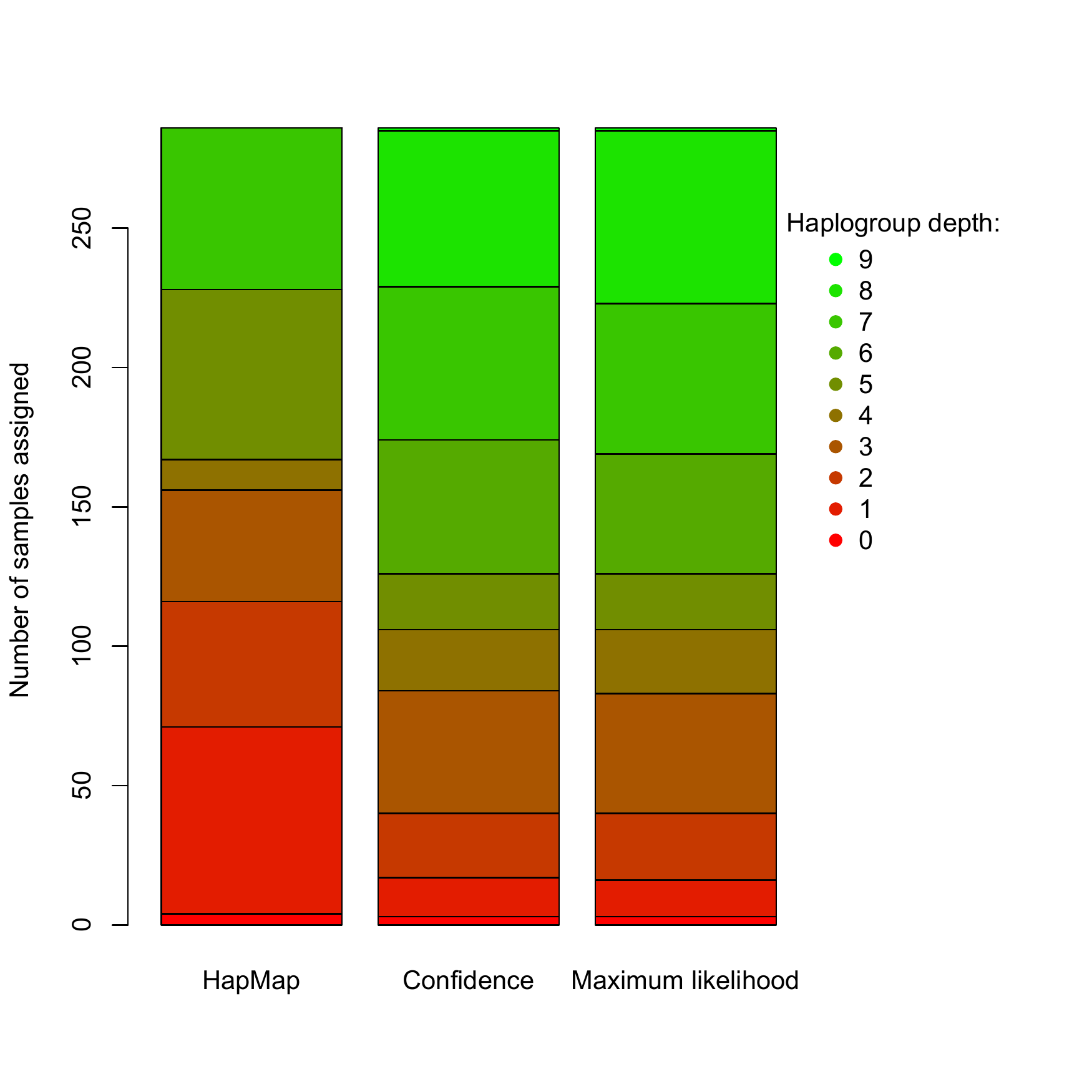}
 \label{justHM}
 \caption{The distribution of haplogroup depth across the manually
   assignment samples using HapMap data, and the confidence and
   maximum likelihood assignments from the low coverage data. A haplogroup depth of
 zero represents samples that cannot be assigned to any major
 haplogroup, a depth of one represents assignment to a major
 haplogroup but no further resolution, and each additional assignment
 adds one to the haplogroup depth.}
\end{figure}

\section{Discussion}

We have presented an efficient statistical method for assigning
haplogroups by maximum likelihood, and shown that it can accurately
and automatically assign haplogroups to short-read data. 

YFitter can be downloaded from
http://sourceforge.net/projects/yfitter/. The software can either be
using in conjunction with the short-read utility program samtools\citep{li2009} to
assign haplogroups to sequencing data, or with the genotype utility
program PLINK\citep{purcell2007} to assign haplogroups to genotype data.

\bibliographystyle{plain}
\bibliography{refs.bib}

\end{document}